\documentclass[aps,preprint,floatfix]{revtex4}

\usepackage{dcolumn}
\usepackage{bm}
\usepackage{graphicx}
\graphicspath{{Images/}}
\usepackage{epstopdf}

\begin{document}

\title{A short walk through the physics of neutron stars}
\author{Isaac Vida\~na}                     

\affiliation{Istituto Nazionale di Fisica Nucleare, Sezione di Catania. Dipartimento di Fisica, Universit\`a di Catania, Via Santa Sofia 64, I-95123, Catania, Italy}

\begin{abstract}
In this work we shortly review several aspects of the physics of neutron stars. After the introduction we present a brief historical overview of the idea of neutron stars as well as of the theoretical and observational developments that followed it from the mid 1930s to the present. Then, we review few aspects of their observation discussing, in particular,
the different types of telescopes that are used, the many astrophysical manifestations of these objects, and several observables such as masses, radii or gravitational waves. Finally, we briefly summarize some of theoretical issues like 
their composition, structure equations, equation of state, and neutrino emission and cooling. 

\end{abstract}

\maketitle


\section{Introduction}
\label{sec:intro}

The answer to the question, {\it what is a neutron star ?}, is not unique and depends on who is asked. Astronomers would probably answer that these objects are very little stars observed as radio pulsars or sources of X- and $\gamma$-rays, whereas particle physicists would say that they are neutrino sources (mainly when they are born) and probably  the only places in the Universe where deconfined quark matter may be really abundant. Cosmologists would reply that neutron stars are almost black holes, in the sense that they are very compact objects but not as compact as black holes. Finally, for nuclear physicists they are the biggest neutron-rich nuclei of the Univese with mass numbers of the order $A\sim 10^{56}-10^{57}$, radii of about $10-12$ km, and masses in the range $M \sim 1-2M_\odot$ (being $M_\odot\simeq 2\times 10^{33}$ g the mass of the Sun). Everybody, however, agrees on the fact that 
neutron stars are a type of stellar compact remant that can result from the gravitational collapse of  
an ordinary star with a mass in the range $8-25M_\odot$ during a Type-II, Ib or Ic supernova event. A supernova explosion occurs when an ordinary star has exhausted all its possibilities for energy production by nuclear fusion. When this happens, the pressure gradient provided by radiation is not sufficient to balance the gravitational attraction, consequently, the star becomes unstable and it eventually collapses. The inner dense regions of the star collapse first and gravitational energy is released and transferred to the outer layers, blowing them away. After the supernova only a fraction of the star is left, and this final product might be a white dwarf, a neutron star (the topic of this lecture) or a black hole, depending on the initial mass of the progenitor star.

Neutron stars are excellent observatories to test our present knowledge of the fundamental properties of matter 
under the influence of strong gravitational and magnetic fields at extreme conditions of density, isospin asymmetry and temperature. They offer an interesting interplay between nuclear processes and astrophysical observables. Their study constitutes nowadays one of the most fascinating  fields of research that requires expertise from different disciplines like general relativity, high-energy physics, nuclear and hadronic physics, neutrino physics, quantum chromodynamics (QCD), superfluid hydrodynamics, plasma physics or even solid state physics. Enormous theoretical advances has been done in understanding the extreme and unique properties (magnetic fields, rotational frequencies, gravitational fields, surface temperatures, ... ) of these exotic objects. Major advances have been also achieved in their observation. The new generation of space X-ray and $\gamma$-ray observatories are enabling new observations and breakthrough discoveries ({\it e.g., }kHz quasi-periodic oscillations, bursting millisecond pulsars, half-day long X-ray superbursts). The  thermal emission from isolated neutron stars provides important information on their cooling history and it allows the determination of their radii. At the same time, improvements in radio telescopes and interferometric techniques have increased the number of known binary pulsars, allowing for extremely precise neutron star mass  measurements and tests of general relativity. A large multinational effort has taken place in the last decade to build a new generation of graviational wave detectors which has been recenlty rewarded with the exciting observation of the first signal from the merger of two neutron stars \cite{gw}.

This work is the result of a lecture given at the school {\it ``Rewriting Nuclear Physics Textbooks: Basic nuclear interactions and their applications to nuclear processes in the Cosmos and on Earth"}, held at the University of Pisa during the last week
of July 2017. To review in a complete and detailed way all the physics of neutron stars in the time of one hour and a half of this lecture is basically an impossible task and, therefore, our scope here is just simply to present a brush-stroke on this topic. The interested reader can find several excellent books and many reviews that comprehensively cover all different aspects of this facinating field \cite{shapiro,weber,glendenning,hpy}. 

The manuscript follows the scheme of the lecture and it is organized in the following way. A brief historical introduction on the idea of neutron stars and the further theoretical and observational developments that followed this idea is presented in Sec.\ \ref{sec:history}. In Sec.\ \ref{sec:observation} we discuss different aspects on the observation of neutron stars whereas the theoretical ones are reviewed in Sec.\ \ref{sec:nstheory}. Finally, a summary and few concluding remarks are shortly presented in Sec.\ \ref{sec:summary}.


\section{Brief historical overview}
\label{sec:history}

The possible existence of neutron stars was proposed by Baade and Zwicky \cite{ba34} in 1934 only two years after the discovery of the neutron by Chadwick \cite{ch32}, although it has been suggested (see {\it e.g.} Refs.\ \cite{hpy,baym82} for details of this story)
that Landau, already in 1931, speculated about the possible existence of stars more compact than white dwarfs containing very dense matter \cite{landau32}. Baade and Zwicky pointed out that a massive object consisting mainly of neutrons at very high density would be much more gravitationally bound than ordinary stars. They also suggested that such objects could be formed in supernova explosions. In 1939, Tolman \cite{to39} and, independently,  Oppenheimer and Volkoff \cite{op39} derived the equations that describe the structure of a static star with spherical symmetry in general relativity, and performed the first theoretical calculation of the equilibrium conditions of neutron stars and their properties assuming an ideal gas of free neutrons at high density. As a curiosity, it is interesting to mention that in 1934 Chandrasekhar and von Neumann obtained the same equations of hydrostatic equilibrium, although they did not publish their work (see again Refs.\ \cite{hpy,baym82} for more details). In their calculation, Oppenheimer and Volkoff found that stable static neutron stars could not have masses larger than $\sim 0.7M_\odot$, a value much lower than the Chandrasekhar mass limit of white dwarfs $\sim 1.44M_\odot$. 

The low value of the maximum mass $M_{max}$ obtained by Oppenheimer and Volkoff is consequence of the simple description they made of the state of matter of the neutron star interior in terms of non-interacting neutrons, and it is an indication that the role of the nuclear forces is fundamental to determine properly the structure of these objects. Further progress towards the construction of a more realistic neutron star matter equation of state (EoS) was done after the Second World War. In the mid-1950s Wheeler and collaborators \cite{harrison58} calculated the EoS of a non-interacting neutron-proton-electron gas under the conditions of $\beta$-equilibrium ({\it i.e.,} in equilibrium with respect to weak interaction processes).  In 1959 Cameron \cite{cameron59} used Skyrme-type forces \cite{skyrme59} to study for the first time the effect of the nucleon-nucleon interaction on the EoS and structure of neutron stars finding values of $M_{max} \sim 2M_\odot$. At that time it started to be clear that, in addition to neutrons, protons and electrons, other particles such as muons, mesons or hyperons (baryons with strangeness content) could also be present in the interior of neutron stars. The effects of hyperons in neutron stars were discussed qualitatively first by Cameron \cite{cameron59} and Salpeter \cite{salpeter60}. These works were followed soon after by the first detail calculations of Ambartsumyan and Saakyan \cite{as60} of the EoS and composition of an equilibrated mixture of degenerate free Fermi gases of baryons, mesons and leptons. Few years after, Tsuruta and Cameron \cite{tscameron66} analized the role of baryon-baryon forces on the dense matter EoS with hyperons using phenomenological interactions. The possibility of pion condensation in dense nuclear matter
and its influence on the properties of neutron stars was pointed out by Migdal \cite{migdal71} and, independently, by Sawyer \cite{sawyer72} and Sawyer and Scalapino \cite{sawyerb72} at the beginning of the 1970s, and then considered by many other authors. Kaon condensation has also been extensively considered in the literature since the pubication of the work of Kaplan and Nelson \cite{kapnel86} in 1986.
The possible existence of deconfined quark matter in the core of neutron stars was already considered in 1965 by Ivanenko and Kurdgelaidze \cite{iva65,iva69} about one year after the introduction of the quark model by Gell-Mann \cite{gellmann64} and Zweig \cite{zweig64}. Many authors have followed the footprints of all these pioneers and, although, major advances have been achieved on our understanding of the properties of matter under the extreme conditions found in the interior of neutron stars, the true behaviour of the nuclear EoS at very high densities remains still uncertain. The interested reader is referred to Refs.\ \cite{oertel16} and \cite{fiorella18} for two recent excellent reviews on the nuclear EoS, and to Ref.\ \cite{epja} for a comprehensive monograph on the role of exotic degrees of freedom in neutron stars.

Another important theoretical step was done in 1959 when Migdal \cite{migdal59} suggested that neutron superfluidity could occur in neutron star interiors. In 1964 Ginzburg and Kirzhnits \cite{ginz64} estimated the energy gap $\Delta$ of a pair of neutrons in a singlet-state at densities $\rho=10^{13}-10^{15}$ g/cm$^3$ finding values in the range $5-20$ MeV. Wolf \cite{wolf66} made a very important step by showing that the singlet-state neutron pairing operates at subnuclear densities in the inner crust of the neutron star but disapears in the core where the neutron-neutron interaction becomes repulsive. The possibility of having in the core proton pairing in a singlet-state and neutron pairing in triplet-state was understood later. Since then, superfluidity in nuclear matter has received a great deal of attention due to its important consequences for a number of neutron star phenonema such as, pulsar glitches or cooling. 

Most of the theoretical effort in the 1960s was focused on modeling neutron star cooling motivated by the expectations  of detecting the thermal emission from their surfaces. First estimates were done by Stabler \cite{stabler60} in 1960 followed four years later by Chiu \cite{chiu64} who repeated the estimates and proved theoretically the possibility to discover neutron stars from their thermal emission. First simple neutron star cooling calculations were performed by Morton \cite{morton64}, Chiu and Salpeter \cite{chiusal}, and Bahcall and Wolf \cite{bahcalla,bahcallb}. The latter authors pointed out that cooling rates depend strongly on the neutrino emission processes and that this dependence can be used to constraint the dense matter EoS by comparing theoretical cooling models with the observation of the thermal radiation from neutron stars. The main elements of a strict neutron star cooling theory, such as the neutrino and photon cooling stages or the relation between the internal and surface temperatures, were formulated by Tsuruta and Cameron \cite{tsuruta66} in 1966. The reader can find more details of the history of neutron star cooling in the work by Yakovlev {\it et al.} \cite{yakovlev99}.
  
Despite the theoretical effort, neutron stars were ignored by the astronomical community for about 30 years since their existence was theoretically hypothesized in the mid 1930s. A reason often given to neglect the neutron star idea was that because of their small area, their residual thermal radiation would be too faint to be observed at astronomical distances with optical telescopes in comparison with ordinary stars. However,  the discovery of the first cosmic X-ray source of non-solar origin, Sco X-1, in rocket experiments by Giacconi {\it et al.} \cite{giacconi62} in 1962 generated a great interest in neutron stars. First attempts to prove these newly discovered compact X-ray sources and neutron stars, nevertheless, failed. Different methods to discover neutron stars were poposed. Zeldovich and Guseynov \cite{zeldovich66}, for instance, proposed to observe some selected binaries with optical primary components and invisible secondary components, assuming that the latter are neutron stars. In 1967 Pacini \cite{pacini67} showed that 
a rapidly rotationg neutron star with a strong dipole magnetic field could transform its rotational energy into electromagnetic radiation and accelerate particles to high energies, powering in this way a surrounding nebula, like the Crab nebula.

On August 6$^{th}$ 1967 the first radio pulsar, named PSR B1919+21\footnote{Pulsars are designated with the prefix ``PSR" followed by the pulsar right ascension and the degrees of declination. The right ascension is also prefixed with a ``J" (Julian epoch) or a ``B" (Besselian epoch).}, was discovered by Bell and Hewish \cite{be68}. 
They identified a 81.5 Mhz source with a pulsating period of 1.377 s. The identification of pulsars with neutron stars, however, was not immediately obvious to most astrophysicists. The first argument that observed pulsars were in fact rotating neutron stars with strong surface magnetic fields of the order of $\sim 10^{12}$ G was put forward by Gold \cite{go68}. He pointed out that such objects could explain many of the observed features of pulsars, such as {\it e.g.,} the remarkable stability of the pulse period. Gold predicted a small increase in the period as rotational enery is lost due to radiation. Showly after, this was confirmed when a slowdown of the Crab pulsar was discovered. Because of this success and the failure of other models, pulsars were (and are) thought to be highly magnetized rotating neutron stars.

Since 1968, there has been much theoretical work to understand the  properties of neutron stars. This was further stimulated by the discovery of pulsating compact X-ray sources (``X-ray pulsars") by the UHURU satellite in 1971. 
These sources are believed to come from a neutron star in a close binary system which is accreting matter from its ordinary companion star. The evidence for the formation of neutron stars in supernova explosions was provided by the simultaneous discoveries of the Crab and the Vela pulsar in the late fall of 1968, both of which are located in supernova remnants, confirming the prediction of Baade and Zwicky. The Crab nebula, for instance, is in fact the remnant of the historical supernova explosion observed by Chinese astronomers in 1054 A.D.  

A further step in the history of neutron star observation was done in 1974 when Hulse and Taylor \cite{hulsetaylor} discovered the first binary pulsar PSR J1913+16, known popularly simply as the Hulse--Taylor pulsar. This system is formed by two neutron stars orbiting around their common center of mass.  During the 1980s, 1990s and 2000s several satellites with on board X-ray and $\gamma$-ray telescopes devoted to the observation of neutron stars have been launched and more will be launched in the future. 

Neutron star history still reserves us many surprises, being the last one the very recent first direct detection of gravitational waves from the merger of two neutron stars \cite{gw}, exactly 50 years and 11 days after the discovery of the first radio pulsar.


\section{Observation of neutron stars}
\label{sec:observation}

\begin{figure}[t!]
\begin{center}
\includegraphics[width=12cm]{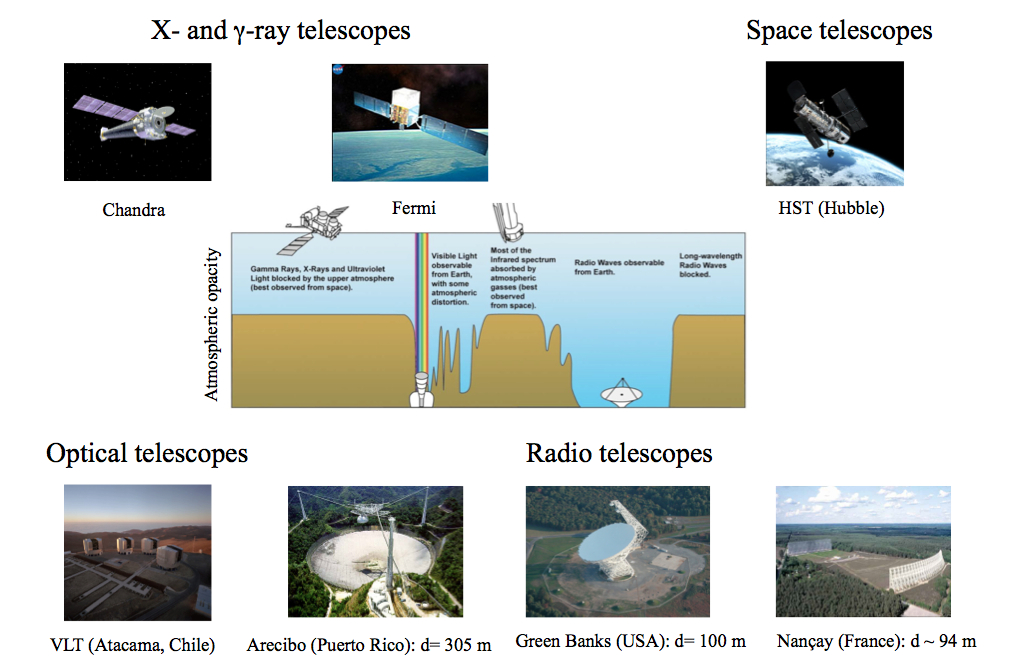} 
\end{center}
\caption{(Color online). Different types of ground-based and on-board telescopes covering all band of the electromagnetic spectrum used for the observation of neutron stars. Figures of the telescopes have been taken and adapted from the web.} 
\label{f:obs}    
\end{figure}

\subsection{Telescopes for neutron star observations}
\label{sec:telescopes}

Neutron stars are observed in all bands of the electromagnetic spectrum: radio, infrared, optical, ultraviolet, X-ray and $\gamma$-ray. Their observation requires different types of ground-based and on-board telescopes such as those shown as example in Fig.\ \ref{f:obs}. Radio observations are carried out with ground-based antennas located in different places of the world like, for instance, the {\it Arecibo radio telescope} in Puerto Rico, the {\it Green Bank Observatory} in West Virginia, or the  
{\it Nan\c{c}ay decimetric radio telescope} in France. Large ground-based telescopes like the {\it Very Large Telescope} (VLT) in the Atacama desert in Chile can be used to perform observations in the near infrared and the optical bands. Ultraviolet and optical observations can be also performed with the help of the {\it Hubble-Space Telescope} (HST).
Observations in the extreme ultraviolet, X-ray and $\gamma$-ray require the use of space observatories. Examples of
these observatories are: the {\it Chandra X-ray Observatory} (CXO), the {\it X-ray Multi Mirror} (XMM-Newton) and the {\it Rossi X-ray Timing Explorer} (RXTE) in the case of X-ray observations; and the {\it High Energy Transient Explorer} (HETE-2), the {\it International Gamma-Ray Astrophysics Laboratory} (INTEGRAL) and the {\it Fermi Gamma-ray Space Telescope} (FGST), in the case of $\gamma$-ray ones.

Information on the properties of neutron stars can be obtained not only from the observation of their electromagnetic radiation but also through the detection of the neutrinos emitted during the supernova explosion that signals the birth of the star. Some examples of past, present and future neutrino observatories around the world are: the under-ice telescopes AMANDA ({\it Antarctic Muon And Neutrino Detector Array}) and its succesor the {\it IceCube} observatory
both placed in the South Pole; the under-water projects ANTARES ({\it Astronomy with a Neutrino Telescope and Abyss environmental RESearch}) and the future KM3NET ({\it Cubic Kilometre Neutrino Telescope}) in the Mediterranean sea; or the underground observatories SNO ({\it Sudbury Neutrino Observatory}) 
located 2100 meters underground in the Vale's Creighton Mine in Canada, and the {\it Kamioka} observatory placed at the Mozumi Mine near the city of Hida in Japan.

Gravitational waves offer a new way of observing neutron stars that requires the help of the new generation of gravitational observatories such as the ground-based Advanced LIGO ({\it Laser Interfrometer Gravitational-Wave Observatory}) and Advanced VIRGO (from the {\it European Gravitational Observatory}), or the future European Space Agency mission LISA ({\it Laser Interferometer Space Antenna}) planned to be launched approximately in 2034.


\subsection{The one thousand and one observational faces of neutron stars}
\label{sec:1001faces}

Neutron stars can be observed either as isolated objects or forming binary systems together with other neutron stars, white dwarfs or ordinary (main-sequence and red giant) stars. Modern theories of binary evolution predict also the existence of binary systems formed by neutron stars and black holes, although this kind of systems have not been discovered yet. In the next, we briefly summarize the very rich observational diversity of neutron stars which includes a large variety of different classes of objects. The interested reader will find, {\it e.g.,} in Ref.\ \cite{hpy} a detailed description of these objects and their phenomenology. 

Isolated neutron stars are mostly detected as {\it radio pulsars} although they have been also observed in other frequencies. A pulsar ({\it radio pulsar}, {\it X-ray pulsar} or {\it $\gamma$-ray pulsar}, depending on the spectral range in which pulsations are observed) is a highly magnetized rotating neutron star that emits a beam of electromagnetic radiation which is observed as a pulse only when the beam is pointing towards the Earth. Depending on the particular mechanisms which power their activity pulsars can be classified as {\it rotational powered}, {\it accretion powered} or {\it magnetically powered} pulsars. A simple model of a pulsar will be presented later in this article. 

Not all isolated neutron stars, however, pulsate. A class of non-pulsating objects are the so-called {\it radio-quite isolated neutron stars}. Two interesting types of them are the {\it compact central objects} (CCOs) in supernova remnants and the {\it dim isolated neutron stars} (DINS) not associated directly with supernova remnants. The CCOs are point-like young  X-ray souces located near the center of supernova remnants with unusually weak magnetic fields. The DINS are neutron stars with an age between $10^5-10^6$ years characterized by pure thermal blackbody X-ray spectra with effective temperatures $\sim$ ($0.5-1$)$\times 10^6$ K. They are also called {\it X-ray thermal neutron stars}. 

Other types of isolated neutron stars are the {\it soft gamma repeaters} (SGRs) and the {\it anomalous X-ray pulsars} (AXPs). SGRs are sources soft $\gamma$-ray and X-ray bursts which repeat at irregular intervals. The duration of a typical burst is $\sim$ 0.1 s and its energy $\sim 10^{41}$ erg. It is thought that SGRs could be a type of {\it magnetar} (a young, isolated highly magnetized neutron star) or, alternatively, neutron stars with fossil disks around them. AXPs are sources of pulsed X-ray emission with pulsation periods ranging from 6 to 12 s and X-ray luminosities from $\sim  10^{33}$ to $10^{35}$ erg/s. These pulsars differ from the classical X-ray pulsars in X-ray binaries (see below) by the absence of any evidence that they form a binary system. AXPs share many similarities with SGRs like the existence of bursting activity also in this class of objects. This has motivated the identification of AXPs also with {\it magnetars}.  Currently it is assumed that AXPs and SGRs belong in fact to the same class of neutron stars and, in particuar, SGRs are thought to be younger than AXPs and to transform into them during the course of their evolution.

Neutron star binaries can be wide systems, in which there is no mass exchange between the two objects forming it, or more compact systems where mass is transferred to the neutron star from its companion. In the first case neutron stars behave usually as isolated objects, whereas in the second one binaries are mainly observed as X-ray sources. Such systems are called {\it X-ray binaries} and can be observed as {\it X-ray pulsars}, {\it X-ray bursters} or sources of {\it quasiperiodic X-ray oscillations}. The X-ray emission in these systems is produced near the neutron star surface and/or in the accreation disks. {\it X-ray binaries} are usually classified as {\it high-mass X-ray binaries} (HMXBs) if the mass of the companion object is $M_{c} \geq (2-3)M_\odot$, and {\it low-mas X ray binaries} (LMXRs) when
$M_{c}\leq M_\odot$. Neutron stars companions in HMXRs are usually O-B stars, while in LMXRs they are dwarfs stars, particularly red dwarfs. Depending on the regularity or irregulariry of their activity {\it X-ray binaries} can also be classified as {\it persistent} or {\it transient} sources. The latter, called {\it X-ray transients}, are sources which go from active to a quiescent states and back on timescales of some hours and longer.  {\it X-ray binaries} show a very rich and complicated phenomenology which is a reflection of the complex nature of these soures that is still far from being completely understood.

Let us finally mention that neutron stars can also host {\it exoplanets} (also called {\it extrasolar planets}, {\it i.e.,} planets outside the solar system). The first example of this kind of system is the pulsar PSR B1257+12 in the constellation of Virgo which has a planetary system formed by three {\it exoplanets} named ``Draugh" (PSR B1257+12b or PSR B1257+12A), ``Poltergeist" (PSR B1257+12c or PSR B1257+12B) and ``Phobetor" (PSR B1257+12d or PSR B1257+13C) discovered in 1992 (Poltergeist and Phobertor) and 1994 (Draugh). The second example is the neutron star-white dwarf binary system PSR B1620-26 in the constellation of Scorpius which, in 2000, was confirmed to host an {\it exoplanet} orbiting the two stars. There are other candidates which, however, have not been confirmed yet.


\subsection{Neutron star observables}

After fifty years of observations we have collected an enourmous amount of data on different neutron star observables that include: masses, radii, rotational periods, surface temperatures, gravitational redshifts, quasiperiodic oscillations, magnetic fields, glitches, timing noise and, very recently, gravitational waves. In the following lines we shortly review these observables.

\subsubsection{Masses}

Neutron star masses can be inferred directly from observations of binary systems and likely also from supernova explosions. There are five orbital (or Keplerian) parameters which can be precisely measured in any binary system. These are: the orbital period ($P_b$), the projection of the pulsar's semimajor axis on the line of sight ($x\equiv a_1 \mbox{sin}\, i/c$, where $i$ is inclination of the orbit), the eccentricity of the orbit ($e$), and the time ($T_0$) and longitude ($\omega_0$) of the periastron. Using Kepler's Third Law, these parameters can be related to the masses of the neutron star ($M_p$) and its companion ($M_c$) though the so-called mass function
\begin{equation}
f(M_p,M_c,i)=\frac{(M_c\,\mbox{sin}\,i)^3}{(M_p+M_c)^2}=\frac{P_bv_1^3}{2\pi G}
\label{eq:massfun}
\end{equation}
where $v_1=2\pi a_1\mbox{sin}\,i/P_b$ is the projection of the orbital velocity of the neutron star along the line of sight. If only one mass function can be measured for a binary system, then one cannot proceed further than Eq.\ (\ref{eq:massfun}) without additional assumptions. Fortunatelly, deviations from the Keplerian orbit due to general relativity effects can be detected. These relativistic corrections are parametrized in terms of one or more post-Keplerian  parameters. The most significant ones are: the advance of the periastron of the orbit ($\dot \omega$), the combined effect of variations in the transverse Doppler shift and gravitational redshift around an elliptical orbit ($\gamma$), the orbital decay due to the emission of quadrupole gravitational radiation ($\dot P_b$), and the range (r) and shape (s) parameters that characterizes the Shapiro time delay of the pulsar signal as it propagates through the gravitational field of its companion. These post-Kepletian parameters can be written in terms of measured quantities and the masses of the star and its companion as \cite{taylor92}:
\begin{equation}
\dot\omega=3n^{5/3}T_\odot^{2/3}\frac{(M_p+M_c)^{2/3}}{1-e} \ ,
\label{ec:ome}
\end{equation}
\begin{equation}
\gamma=eT_\odot^{2/3}\frac{M_c(M_p+2M_c)}{n^{1/3}(M_p+M_c)^{4/3}} \ ,
\label{ec:gam}
\end{equation}
\begin{equation}
\dot P_b=-\frac{192\pi}{5}(nT_\odot)^{5/3}\left(1+\frac{73}{24}e^2+\frac{37}{96}e^4\right)
\frac{1}{(1-e^2)^{7/2}}\frac{M_pM_c }{(M_p+M_c)^{1/3}} \ ,
\label{ec:pbdot}
\end{equation}
\begin{equation}
r=T_\odot M_c \ ,
\label{ec:r}
\end{equation}
\begin{equation}
s=x\frac{n^{2/3}}{T_\odot^{1/3}}\frac{(M_p+M_c)^{2/3}}{M_c} \ ,
\label{ec:s}
\end{equation}
where $n=2\pi/P_b$ is the orbital angular frequency and $T_\odot\equiv GM_\odot/c^3=4.925490947\times 10^{-6}$s. The measurement of any two of these post-Keplerian parameters together with mass function $f$ is
sufficient to determine uniquely the masses of the two components of the system. An example of a high precision mass measurement is that of the famous Hulse--Taylor binary pulsar \cite{hulsetaylor} with measured masses
$M_p=1.4408\pm 0.0003 M_\odot$ and $M_c=1.3873 \pm 0.0003 M_\odot$. Another examples are those of the recently observed millisecond pulsars
PSR J1614-2230 \cite{demorest} and PSR J0348+0432 \cite{antoniadis} with masses $M_p=1.928\pm 0.017 M_\odot$ and $M_p=2.01\pm0 0.04 M_\odot$, respectively. These are binary systems formed by a neutron star and white dwarf.

\subsubsection{Radii}

Neutron star radii are very difficult to measure mainly because neutron stars are very small objects and are very far away from us ({\it e.g.,} the closest neutron star is probably the object RX J1856.5-3754 which is about 400 light-years from Earth). Direct measurements of radii do not exist. However, a possible way to determine them is to use the thermal emission of low-mass X-ray binaries. The observed X-ray flux ($F$) and temperature ($T$), assumed to be originated from a uniform blackbody, together with a determination of the distance ($D$) of the star can be used to obtained an effective radius 
\begin{equation}
R_\infty=\sqrt{\frac{FD^2}{\sigma T^4}} \ .
\label{ec:radinf}
\end{equation}
Here $\sigma$ is the Stefan--Boltzmann constant. The neutron star radius $R$ can be then obtained from  $R_\infty$ through the equation
\begin{equation}
R=R_\infty\sqrt{1-\frac{2GM}{c^2R}} \ ,
\label{ec:radns}
\end{equation}
where $M$ is the mass of the star. The major uncertainties in the measurement of the radius throug Eqs.\ (\ref{ec:radinf})-(\ref{ec:radns}) come from the determination of the temperature (see Sec.\ \ref{sec:temp}), which requires the assumption of an atmospheric model, and the estimation of the distance of the star. The analysis of present observations from quiescent low-mass X-ray binaries is still controversial. Whereas the analyis of Steiner {\it et al.} \cite{steiner1,steiner2} indicates neutron star radii in the range of $10.4-12.9$ km, that of Guillot {\it et al.} \cite{guillot1,guillot2} points towards smaller radii of $\sim 10$ km or less. If the result of Guillot {\it et al.} is confirmed by further analysis then the symultaneous existance of massive neutron stars like {\it e.g.} PSR J1614-2230 and PSR J0348+0432 and objects with small radii would be a very complicated problem to solve for any of the existing models for pure nucleonic EoS. A solution to this problem could be the possible existence of the so-called ``twin stars", stars composed of strange hadronic or quark matter \cite{drago14} with similar masses but smaller radii than those made only of nucleons.


\begin{figure}[t!]
\begin{center}
\includegraphics[width=12cm]{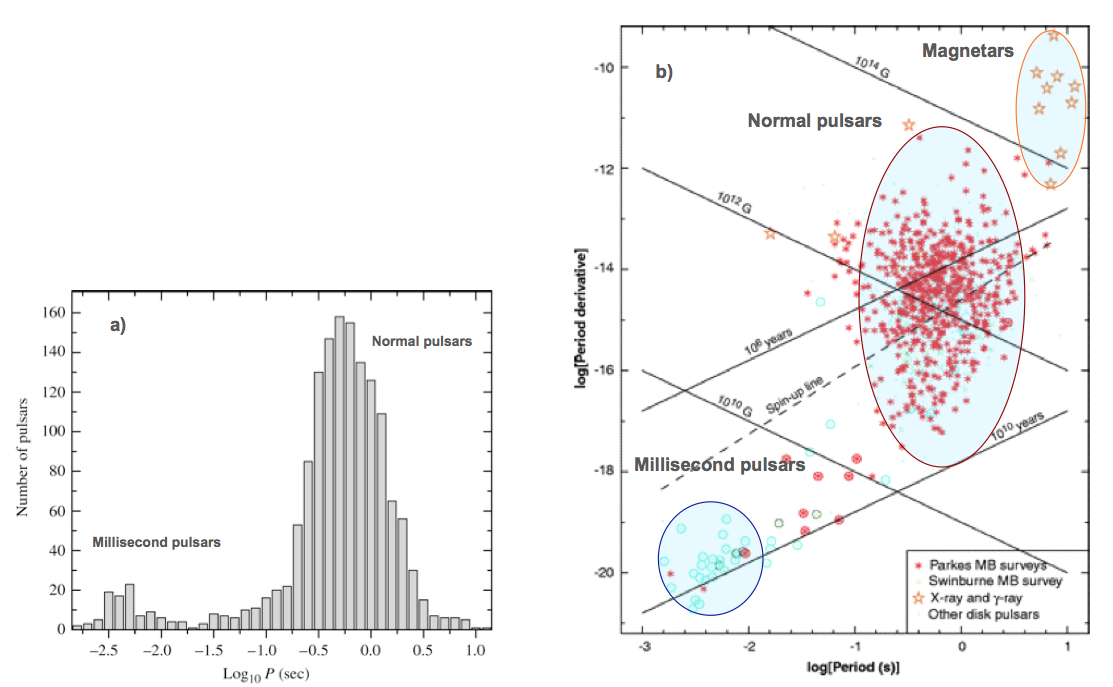} 
\end{center}
\caption{(Color online) Pulsar rotational period distribution (a) and $P\dot P$ diagram (b). The two classes of normal and millisecond pulsars are clearly seen in the period distribution.} 
\label{f:spinp}       
\end{figure}

\subsubsection{Rotational periods}

The major part of the known neutron stars are observed as radio pulsars many of which exhibit very stable rotational periods.
Thanks to highly accurate pulsar timing, observers have been able to measure the rotational period $P$ and the first time derivative $\dot P$ and sometimes even the second one $\ddot P$ of many radio pulsars (see {\it e.g.} \cite{lyne98,lorimer01,livingstone06} and references therein). The distribution of the observed pulsar rotational periods (see Fig.\ \ref{f:spinp}a) shows two clear peaks that indicate the existence of two different classes of pulsars: that of the {\it normal} pulsars with rotational periods of the order of $\sim 1$s, and that of the so-called {\it millisecond} pulsars with rotational periods three orders of magnitude smaller. The first millisecond pulsar was discovered in 1982 with the help of the Arecibo radio telescope and nowadays more than 200 pulsars of this class are known. At present the fastest pulsar known till now, with a rotational period of 1.39595482ms, is the object named PSR J1748-2446ad which was discovered in 2005 in the globular cluster Terzan 5 in the Sagittarius constellation. 

Observed periods $P$ and their first time derivatives $\dot P$ are usually plotted in the so-called {\it $P\dot P$ diagram} (see Fig.\ \ref{f:spinp}b). It is an evolutionary diagram that allows to follow the live of pulsar and to estimate, among other parameters, the pulsar age or its magnetic field strength $B$ (see Sec.\ \ref{sec:pulsmod}). The $P\dot P$ diagram plays a role similar to that of  Hertzsprung--Russell diagram for ordinary stars.

\subsubsection{Surface temperatures}
\label{sec:temp}

The detection of thermal photons from the stellar surface in X-ray binaries allows to determine effective surface temperatures of neutron stars by fitting the observed spectra to blackbody ones. However, one should keep in mind that neutron stars are not blackbodies, because the hydrogen and helium (or even carbon) in their atmospheres (see Sec.\ \ref{sec:anatomy}) modifies the blackbody spectrum. In addition the presence of strong magnetic fields can also modify the surface emission. Surface temperatures are in fact reduced when realistic atmosphere models are used in the fit of the measured spectrum. We note that any uncertainty in the determination of the surface temperature changes the corresponding luminosity $L$ of the star by a large factor according to the Stefan--Boltzmann law, $L=4\pi R^2\sigma T^4$. Therefore, it is not appropiate to use the surface temperature when comparing with observational data but the luminosity instead.

\subsubsection{Gravitational redshift}

An important source of information on the structure of a neutron star is provided by the measurement of its gravitational redshift
\begin{equation}
z=\left(1-\frac{2GM}{c^2R}\right)^{-1/2}-1 
\label{ec:z}
\end{equation}
which allows to contrain the $M/R$ ratio. The interpretation of measured $\gamma$-ray bursts, as gravitationally redshifted 511 KeV $e^{\pm}$ annihilation lines from the surface of neutron stars, supports a neutron star redshift range of $0.2 \leq z \leq 0.5$ with the highest concentration  in the narrower range $0.25 \leq z \leq 0.35$ \cite{liang86}.

\subsubsection{Quasiperiodic oscillations}

Quasiperiodic X-ray oscillations (QPOs) in X-ray binaries measure of the difference between the rotational frequency of the neutron star and the Keplerian frequency of the innermost stable orbit of matter elements in the accretion disk formed by the diffused material of the companion in orbital motion around the star. Their observation and analysis can provide very useful information to understand better the innermost regions of accretion disks as well as to put stringent constraints on the masses, radii and rotational periods of neutron stars. QPOs may also serve as unique proves of strong field general relativity. However, the theoretical interpretation of QPOs is not simple and remains still controversial.

\subsubsection{Magnetic fields}

Since the suggestion of Gold \cite{go68} pulsars are generally believed to be rapidly rotating neutron stars with strong surface magnetic fields. The strength of the field could be of the order of $10^8-10^9$ G in the case of millisecond pulsars, about $10^{12}$ G in normal pulsars, or even $10^{14}-10^{15}$ G in the so-called {\it magnetars}. 
As its was already mentioned, and it will be briefly explained later in section \ref{sec:pulsmod}, the magnetic field strength of a pulsar can be estimated from the observation of its rotational period $P$ and its first derivative $\dot P$.

Despite the great theoretical effort, there is no general consensus yet regarding the mechanism to generate such a strong magnetic fields in neutron stars. The field could be generated after the formation of the neutron star by some long lived electric currents in the highly conductive neutron star material. Or, it could be simply be a fossil remnant from the progenitor star. In fact, assuming magnetic flux conservation during the birth of a neutorn star, a magnetic field of $\sim 10^{12}$ G could originate from the collapse of a main sequence star with a typical surface magnetic field of $10-10^2$ G. From the nuclear physics point of view, however, the most interesting hypothesis is that the magnetic field could be consequence of a phase transition to a ferromagnetic state in the liquid interior of the neutron star.
This possibility has been examined by many authors (see {\it e.g.,} Refs.\ \cite{vidana1,vidana2} and references therein) using different approaches although the results are still contradictory. Whereas calculations based on phenomenological nuclear interactions predict thes transition to occur at densities $(1-4)\rho_0$ ($\rho_0=2.8\times 10^{14}$ g/cm$^3$ being the normal nuclear saturation density), calculations based on realistic two- and three-nucleon forces exclude it completely.

\subsubsection{Glitches}

Pulsars are observed to spin down gradually  due to the transfer of their rotational energy to the emitted electromagnetic radiation. Sudden jumps $\Delta\Omega$ of the rotational frequency $\Omega$, however, have been observed in several pulsars followed by a slow partial relaxation that can last days, months or years. These jumps, mainly observed from relative young radio pulsars, are known as {\it glitches}. The relative increase of the rotational frequency $\Delta\Omega/\Omega$ vary from $\sim 10^{-10}$ to $\sim 5\times 10^{-6}$. The first glitches were detected from the Crab and Vela pulsars \cite{boy69,rad69,rei69}. Nowadays we know more than 520 glitches in more than 180 pulsars. 

Although the exact origin of glitches is not completely known yet, a promising model \cite{anderson75,alpar77} to explain the observed glitch behavior of pulsars is based on the formation of vortex lines in the neutron superfluid in the inner crust (see Sec.\ \ref{sec:anatomy}) of the neutron star. In this model, glitches are the result of a sudden transfer of angular momentum from the neutron superfluid to the solid crust caused by the unpinning of many vortex lines or by the cracking of the crust to which vortices lines are pinned. Other models proposed to explain the origin of glitches include: starquakes ocurring in the crust and/or the core of the star, magnetospheric instabilities, or instabilities in the motion of the superfluid neutrons.

\subsubsection{Timing noise}

One of the most remarkable properties of radio pulsars is their rotational stability. However, some of them show
slow irregular or quasiregular variations of their pulses over time scales of months, years and longer which have been called {\it pulsar timing noise}. These timing imperfections appear as random walks in the pulsar rotation (with relative variations of the rotational period $\leq 10^{-10}-10^{-8}$), the spindown rate or the pulse phase. Their nature is still
uncertain and many hypotheses have been made (see {\it e.g.} Ref.\ \cite{cordes81}). Careful studies of the pulsar timing noise can provide valuable information on the internal structure of neutron stars.

\subsubsection{Gravitational Waves}

Gravitational waves originated from the oscillation modes of neutron stars or during the coalescence of two neutron stars or a black hole and a neutron star constitute also a valuable source of information. Very recently, on August 17$^{th}$ 2017, the graviational wave signal from a binary neutron star merger was detected for the first time by the Advanced LIGO and Advanced VIRGO collaborations \cite{gw} inaugurating, with the detection of this event (known as GW170817), a new era in the observation of neutron stars.


\section{Neutron star theory}
\label{sec:nstheory}


\subsection{Magnetic dipole model of a pulsar}
\label{sec:pulsmod}

A very basic pulsar model that accounts for many of the observed properties of pulsars is the so-called {\it magnetic dipole model}. The most simple version of this model assumes that the pulsar rotates at a frequency $\Omega$ and possesses a magnetic moment $\vec{\mu}$ oriented at an angle $\alpha$ with respect to the rotation axis. The rotation is assumed to be sufficiently slow so that deviations from the spherical shape of the star can ignored to lowest order. The magnetic moment of a sphere with a pure magnetic dipole field is (see {\it e.g.} Ref.\ \cite{jackson})
\begin{equation}
|\vec{\mu}|=\frac{B_pR^3}{2} 
\label{ec:magm}
\end{equation}
where $B_p$ is the strength of the magnetic field at the pole and $R$ the radius of the sphere (in our case the star). This magnetic dipole radiates energy at a rate
\begin{equation} 
\dot E_{mag} = -\frac{2}{3c^3}|\ddot{\vec{\mu}}|^2
\label{ec:edotmag}
\end{equation}
at expenses of the kinetic rotational energy of the star, {\it i.e.,} $\dot E_{mag}=\dot E_{rot}$. Writing
\begin{equation}
\vec{\mu}=\frac{B_pR^3}{2}\left(\vec{e_{||}}\mbox{cos}\,\alpha
+\vec{e_\bot}\mbox{sin}\, \alpha\,\mbox{cos}\,\Omega t
+\vec{e_\bot'}\mbox{sin}\,\alpha\,\mbox{sin}\,\Omega t
\right)\ ,
\label{ec:magm2}
\end{equation}
where $\vec{e_{||}}$, $\vec{e_\bot}$ and $\vec{e_\bot'}$ are three unitary vectors parallel and orthogonal to the rotation axis, and assuming $\alpha$ and $|\vec{\mu}|$ constant and $\dot\Omega^2 \ll \Omega^4$, one finds
\begin{equation} 
\dot E_{mag}=-\frac{B_p^2R^6\Omega^4\mbox{sin}^2\,\alpha}{6c^3}
\label{ec:edotmag2} \ .
\end{equation}

On the other hand the time derivative of the kinetic rotational energy ($E_{rot}=I\Omega^2/2$), assuming a constant moment of inertia ($\dot I=0$), is 
\begin{equation}
\dot E_{rot}=I\Omega\dot\Omega \ .
\label{ec:erot}
\end{equation}

Equating $\dot E_{mag}$ and $\dot E_{rot}$ one arrives to the so-called {\it pulsar evolution differential equation}
\begin{equation}
\dot \Omega=-K\Omega^3 \,\,\,\,\,\, \mbox{or} \,\,\,\,\,\, P\dot P=(2\pi)^2K
\label{ec:pede}
\end{equation}
with
\begin{equation}
K=\frac{B_p^2R^6\mbox{sin}^2\,\alpha}{6c^3I} \ .
\label{ec:k}
\end{equation}
Note that Eq.\ (\ref{ec:pede}) allows to obtain the strength of the magnetic field in terms observable quantities.
More generally, one can write the pulsar evolution differential equation as
\begin{equation}
\dot \Omega=-K\Omega^n \,\,\,\,\,\, \mbox{or} \,\,\,\,\,\, P^{n-2}\dot P=(2\pi)^{n-1}K
\label{ec:pede2}
\end{equation}
where $n$ is the so-called {\it braking index} which can be obtained by differenciating Eq.\ (\ref{ec:pede2}), assuming $K$ constant, as
\begin{equation}
n=\frac{\Omega\ddot\Omega}{\dot\Omega^2}=2-\frac{P\ddot P}{\dot P^2} \ .
\label{ec:brakind}
\end{equation}
As mentioned in the previous section $P$, $\dot P$ and even $\ddot P$ are observable quantities and, therefore, the braking index can be obtained directly from observation. Deviations from the value $n=3$, which corresponds to the magnetic dipole model, measure the validity of this model. Such deviations can be due, among other things, to a torque acting on the pulsar from outflow particles or to the time dependence of $K$, the monent of intertia $I$ or the angle $\alpha$, which is ignored in this basic model. 

Integrating Eq.\ (\ref{ec:pede2}) from the pulsar birth at $t=0$ to a current age $t$ one obtains
\begin{equation}
t=-\frac{1}{n-1}\frac{P(t)}{\dot P(t)}\left\{1-\left(\frac{P(0)}{P(t)}\right)^{n-1}\right\} \ .
\label{ec:pede4}
\end{equation}
If $P(0) \ll P(t)$ ({\it i.e.,} the newly born pulsar rotates much faster than at the current time) then $t \approx P/((n-1)\dot P)$ which in the case of magnetic dipole moment ($n=3$) reduces to
\begin{equation}
t=\frac{P}{2\dot P}
\label{ec:chage}
\end{equation}
known as {\it pulsar dipole age} or {\it charactetistic pulsar age}. In the case of the Crab pulsar, for instace, the current observed values of $P$ and $\dot P$ are $0.0330847$ s and $4.22765\times 10^{-13}$ s/s, respectively, which give a charactetistic age of about 1240 years, a value in qualitatively good agreement with its true age of 964 years.

Taking the logarithm of Eqs.\ (\ref{ec:pede}) and (\ref{ec:chage}) one gets
\begin{equation}
\mbox{log}\,\dot P=\mbox{log}\left((2\pi^2)K\right)-\mbox{log}\,P \ , 
\label{ec:log}
\end{equation}
\begin{equation}
\mbox{log}\,\dot P = \mbox{log}\,P-\mbox{log}\,(2t) \ .
\label{ec:log}
\end{equation}
These are, respectively, lines of constant magnetic field and characteristic age which plotted in the $P\dot P$ diagram (see Fig.\ \ref{f:spinp}b) allow to estimate the magnetic field and the age of the different objects in the diagram.


\subsection{Are neutron stars made only of neutrons ?}
\label{sec:int_struc}

\begin{figure}[t!]
\begin{center}
\includegraphics[width=12cm]{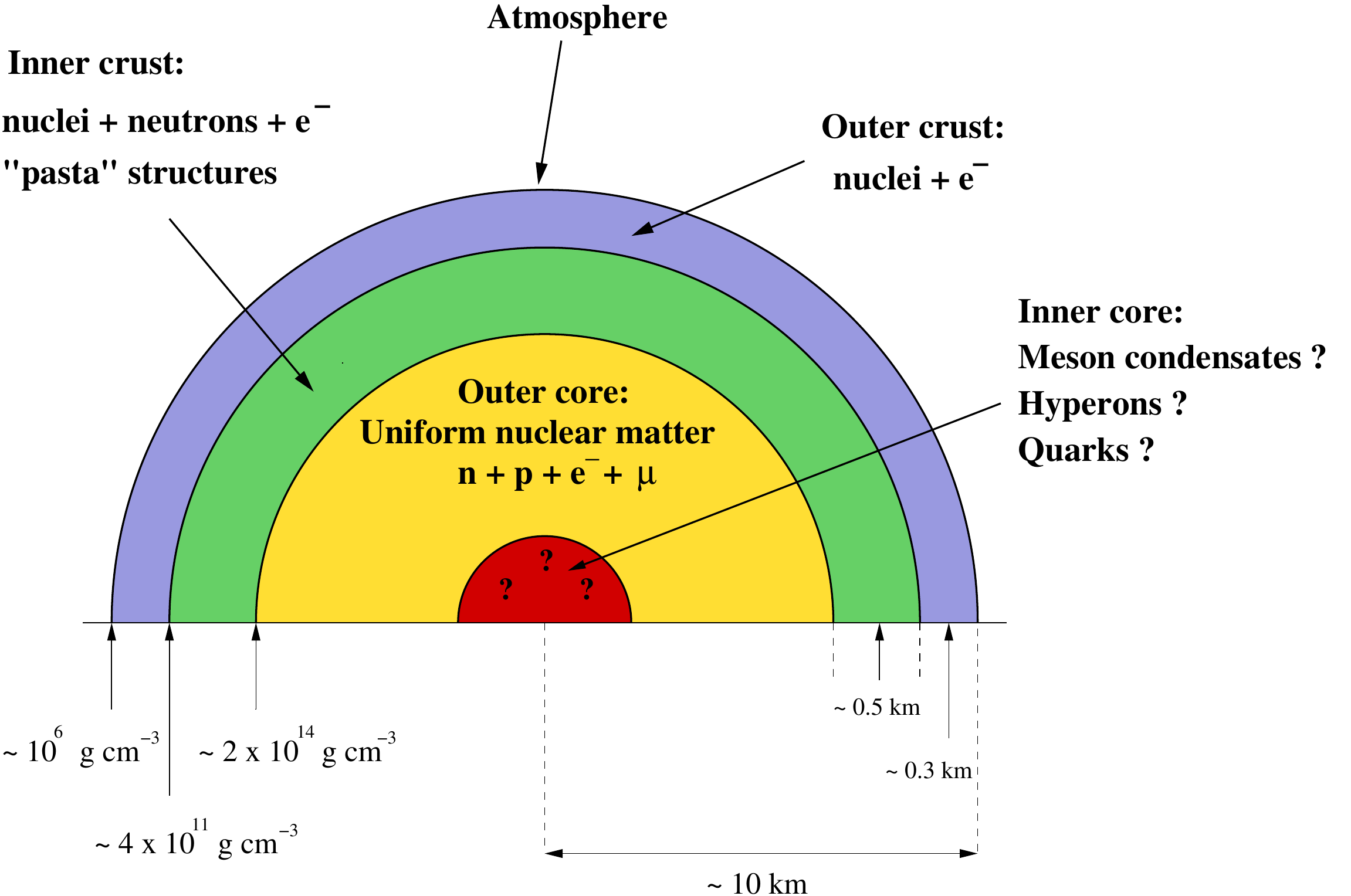} 
\end{center}
\caption{(Color online) A schematic cross section of a neutron star illustrating the various regions discussed in the text. The different regions shown are not drawn on scale.} 
\label{f:star}  
\end{figure}

\subsubsection{Anatomy of a neutron star}
\label{sec:anatomy}

Neutron stars are supported against gravitational collapse mainly by the neutron degeneracy pressure and may have, as already mentioned, typically masses of the order of $1-2M_\odot$ and radii within the range $10-12$ km.  Such masses and radii yield an averaged density for neutron stars of the order of $\sim 10^{14}$ g/cm$^3$. However, the expected densities in neutron stars span a rather wide range, and in fact the internal structure of these objects can be described by an ``onion"-like structure. In Fig.\ \ref{f:star} it is shown a schematic cross section of a neutron star where the reader can distinguish several regions: the {\it atmosphere}, the thin outermost layer; the {\it crust}, divided into {\it outer} and {\it inner crust} with a thickness of about $\sim 1$ km and a mass only a few percent of the total mass of the star; and the {\it core}, divided also into two parts: the {\it outer} and the {\it inner core}, with a radius of about $\sim 10$ km and almost the total mass of the star.

The most external region, the {\it atmosphere}, is a very thin plasma layer where the observed thermal spectrum of the neutron star is formed. Its thickness varies from some ten centimeters in hot neutron stars to a few millimiters in the cold ones. The theoretical study of neutron stars atmospheres has been carried out by many authors (see {\it e.g.,} Ref.\ \cite{zavlin02} and references therein), although current atmosphere models, consisting of hydrogen, helium or carbon, are far from being complete. The main problems are associated with the calculation of the EoS, the ionization equilibrium and the spectral opacity of the atmospheric plasma (see Chapters 2 and 4 of Ref.\ \cite{hpy} for a detailed review).

The {\it outer crust} extends from the bottom of the {\it atmosphere} up to few hundred meters below. It is  a solid region where heavy nuclei, mainly around the iron mass number, form a Coulomb lattice  in $\beta$-equilibrium with a strongly degenerate electron gas which becomes ultrarelativistic at densities $\rho > 10^6$ g/cm$^3$. We note that the study of the mechanical, thermal and electric properties of this region is not only interesting for the physics of neutron stars, but also for solid state physics due to the extreme large values of density, temperature and magnetic fields as compared to laboratory conditions.

Moving towards the interior of the star the increase of the density induces electron capture processes on nuclei,
\begin{equation}
e^- + ^AZ \,\,\, \rightarrow \,\,\, ^A(Z-1) + \nu_e \ ,
\label{e:ecp}
\end{equation}
which become more and more neutron rich. When the density reaches a value $\rho = \rho_{drip} \sim 4\times 10^{11}$ g/cm$^3$ the only available levels for the neutrons are in the continuum and, thus, they start to ``drip out" of the nuclei. The onset of the neutron drip defines the border between the {\it outer crust} and the {\it inner crust}.

The {\it inner crust} can be about one kilometer thick. The density in this region ranges from $\rho_{drip}$ up to $\sim 0.5\rho_0$. Matter here consist of a mixture of very neutron-rich nuclei arranged in a Coulomb lattice, electrons and free neutrons which are expected to be paired in the s-wave by the nuclear residual interaction and, therefore, to form a superfluid. In addition, in this region, the competition between the nuclear and Coulomb forces makes the nuclei to lose their spherical shapes and to adopt more exotic topologies (droplets, rods, cross-rods, salabs, tubes, bubbles shapes) giving rise to what  has been called ``nuclear pasta" phase due to their resemblance with the Italian pasta \cite{raven83}. The observational evidences for the existence of these exotic structures and their potential relevance are not completely clear yet. However, the recent works of Pons {\it et al.,} \cite{pons13} and Newton \cite{newton13} have provided some hints on the possible existence of these structures. Pons {\it et al.,} \cite{pons13} have shown that a highly resistive layer in the innermost part of the crust naturally limits the spin period of isolated X-ray pulsars to a maximum value of 10-20s. This is in agreement with the lack of observation of 
periods longer than 12s for these objects, giving, may be, the first observational evidence for an amorphous {\it inner crust}. Newton \cite{newton13} has shown that the comparison of calculations of the magnetic field decay of neutron stars and their correponding spin evolution with observations suggest a high degree of disorder in the {\it inner crust}, which might also provide evidence for the ``nuclear pasta".

At densities of about $\sim 10^{14}$ g/cm$^3$ the nuclear clusters dissolve into their constituents neutrons and protons, and one enters in the {\it outer core}. This region is a quantum fluid with densities in the range $0.5\rho_0\leq \rho \leq 2\rho_0$ and a thickness of several kilometers. Matter is mainly composed of p-wave superfluid neutrons with a smaller concentration of s-wave superconducting protons and normal electrons and muons which appear as soon as their chemical potential equals that of the electrons ({\it i.e.,} $\mu_\mu=\mu_e$). For low-mass neutron stars, whose central densities are found to be less than $2-3\rho_0$, the {\it outer core} actually constitutes the entire core of the object.

The central densities of the more massive stars could easily reach values up to several times $\rho_0$. In this case, an
{\it inner core} of several kilometers and densities in the range $\rho \geq 2\rho_0$ occupies the central region of the star. Nevertheless, the composition of this region is not well known, and it is still matter of speculation. The different  
hypotheses include: hyperonic matter,  pion or kaon condensates, or deconfined quark matter. In the literature, these additional degrees of freedom are sometimes referred to as {\it exotic}, and their presence in the inner core is simply consequence of the fact that the star lowers its energy with their appearance. The possible existence of deconfined quark matter is particularly interesting becauses it establishes a possible link between neutron stars and QCD, the fundamental theory of the strong interaction.

In the previous paragrahs we have described what one can consider more or less the standard internal structure of a neutron star. However, theoreticians have also speculated about a special type of compact stars whose structure does not correspond with the one just described. These objects are the so-called {\it strange stars} and are thought to be enterely made of a deconfined mixture of up (u), down (d) and strange (s) quarks ({\it strange quark matter}) with perhaps a small fraction of electrons. Their possible existence is a direct consequence of the Bodmer--Witten--Terezawa hypothesis \cite{bodmer71,witten84,terezawa79,terezawa89a,terezawa89b} according to which three-flavour uds quark matter in equilibrium with respect to the weak interactions could be the true ground state of strongly interacting matter rather than $^{56}$Fe ({\it i.e.,} $E_{uds} \leq M$($^{56}$Fe)/56 $\simeq 930$ MeV)).


\subsubsection{Chemical equilibrium in a neutron star}
\label{sec:cheq}

The equilibrium conditions that allow to determine the chemical composition of a neutron star are governed by the weak processes 
\begin{equation}
b_1 \rightarrow b_2 + l + \bar \nu_l \ , \,\,\, b_2 + l \rightarrow b_1 + \nu_l \ ,
\label{eq:wp}
\end{equation}
where $b_1$ and $b_2$ refer to two different types of baryons, $l$ represents a lepton, and $\nu_l$ and $\bar \nu_l$ its respective neutrino and anti-neutrino.

It is generally understood that the knowledge of the particular path that a body or substance may follow in reaching its equilibrium state is needed to determine that state. One possible way to find it consist in minimizing the total energy density of the system constrained by the subsidiary conditions that express the conservation of some components or attributes (hereafter referred to as ``charges") on a timescale longer than a characteristic time of the system. This can be done by using the method of the Lagrange multipliers \cite{mo53}. In the case of a cold neutron star there are only two conserved charges: the total baryonic density and the total electric charge, which is zero (charge neutrality). Strangeness is not conserved on the scale of a star because the timescale of the electro-weak interaction ($\tau \sim 10^{-10}$s) is short in comparison to macroscopic scales. In addition,  in a cold neutron star, neutrinos have diffused out of the star and, therefore, there is no conservation of the leptonic numbers. 

One can then construct a function $F(\rho_{b_1},\rho_{b_2},\cdots,\rho_{b_B},\rho_{l_1},\rho_{l_2},\cdots,\rho_{l_L})$, from the total energy density $\varepsilon(\rho_{b_1},\rho_{b_2},\cdots,$ 
$\rho_{l_1},\rho_{l_2},\cdots,\rho_{l_L})$ and the two constraint equations that express conservation of the total baryonic density and charge neutrality:
\begin{equation}
\begin{array}{c}
F(\rho_{b_1},\rho_{b_2},\cdots,\rho_{b_B},\rho_{l_1},\rho_{l_2},\cdots,\rho_{l_L})= 
\varepsilon(\rho_{b_1},\rho_{b_2},\cdots,\rho_{l_1},\rho_{l_2},\cdots,\rho_{l_L}) \\
\displaystyle{+\alpha\left(\rho_B-\sum_iB_i\rho_{b_i}\right) 
+\beta\left(\sum_iq_{b_i}\rho_{b_i}+\sum_jq_{l_j}\rho_{l_j}\right) }\ .
\label{eq:f}
\end{array}
\end{equation}
The subscripts $b_i$ and $l_i$ run over all types of baryons and leptons, respectively. The quantities $\alpha$ and $\beta$ are the corresponding Lagrange multipliers. The products $q_{b_i}\rho_{b_i}$ and $q_{l_i}\rho_{l_i}$ denote the charge density of those charged particles, with $q_{b_i}$ and $q_{l_i}$ being the corresponding charge numbers, and $B_i$ denoting the corresponding baryonic number.

The minimization condition requires
\begin{equation}
\begin{array}{c}
\displaystyle{\frac{\partial F}{\partial\rho_{b_1}}=0, \ \ ... \ \ , \ \ 
\frac{\partial F}{\partial\rho_{b_B}}=0, 
\frac{\partial F}{\partial\rho_{l_1}}=0, \ \ ... \ \ , \ \
\frac{\partial F}{\partial\rho_{l_L}}=0 \ ,} \nonumber \\ \\
\displaystyle{\frac{\partial F}{\partial\alpha}=0, \ \ \
\frac{\partial F}{\partial\beta}=0 \ .}
\end{array}
\label{eq:NSM.8}
\end{equation}

Remembering that the chemical potential of a species ``$i$'' is just $\mu_i=\partial\varepsilon/\partial\rho_i$,
the above conditions on $F$ yield a set of equations of the type
\begin{equation}
\mu_{b_i}-B_i\alpha+q_{b_i}\beta=0, \ \ \ i=1,...,B \ , 
\label{eq:NSM.9}
\end{equation}
for the baryons. And a set of equations of the type
\begin{equation}
\mu_{l_j}+q_{l_j}\beta=0, \ \ \ j=1,...,L \ ,
\label{eq:NSM.10}
\end{equation} 
for the leptons. 

Eliminating the Lagrange multipliers $\alpha$ and $\beta$, one can obtain a set of relationships among the
chemical potentials. In general there are as many {\it independent} chemical potentials as there are conserved
charges, and all the others can be written in terms of them. As it was said in the case of a neutron star
there are only two conserved charges, and their corresponding chemical potentials are usually assumed to be $\mu_n$
(associated with conservation of the total baryonic density) and $\mu_e$ (associated with charge neutrality). 
Applying then Eqs. (\ref{eq:NSM.9}) and (\ref{eq:NSM.10}) to the neutron and the electron it is found
\begin{equation}
\alpha=\mu_n, \ \ \ \beta=\mu_e  \ ,
\label{eq:NSM.11}
\end{equation}
and replacing (\ref{eq:NSM.11}) on those equations one has that in general the chemical potential of any
particle can be obtained as a linear combination of $\mu_n$ and $\mu_e$, weighted by the baryon and electric
charge carried by the particle:
\begin{equation}
\begin{array}{c}
\mu_{b_i}=B_i\mu_n-q_{b_i}\mu_e, \ \ \ i=1,...,B, \nonumber \\ \\
\mu_{l_j}=-q_{l_j}\mu_e, \ \ \ j=1,...,L \ .
\end{array}
\label{eq:NSM.12}
\end{equation}

There is an alternative way of deriving this set of equations. It consists of writing down all the possible reactions
among the components of matter. One then rewrites the reactions in terms of the different chemical potentials $\mu_i$. If several reactions are possible, there is an equation for each one and the resulting relationships between the chemical potentials allow one to
express the chemical potentials of all the components of matter in terms of the independent ones. 

The solution of this set of equations determines the composition of matter at its ground state for a given density and types of particles. However, it is clear from these equations that not only the weak interaction rules the composition of matter, but also the strong interaction through the explicit value of the chemical potentials.


\subsection{Structure equations of neutron stars}
\label{sec:structure}

\subsubsection{Static Neutron Stars: TOV equations}

A neutron star is one of the densest objects in the Universe, therefore, Einstein's general relativity theory is needed to determine its structure. Einstein's field equations \cite{we72,mi73} for a spherical static star take the form of the familiar Tolman--Oppenheimer--Volkoff (hereafter referred to as TOV) equations \cite{to39,op39} which, using units in which $G=c=1$, read
\begin{equation}
\frac{dp(r)}{dr}=-\frac{\left(\varepsilon(r)+p(r)\right)\left(M(r)+4\pi r^3p(r)\right)}
{r\left(r-2M(r)\right)}
\label{e:tov1}
\end{equation}
and
\begin{equation}
\frac{dM(r)}{dr}=4\pi r^2 \varepsilon(r) \ .
\label{e:tov2}
\end{equation}

By re-writting Eq.\ (\ref{e:tov1}) in the form
\begin{equation}
\frac{dp(r)}{dr}=-\frac{M(r)\varepsilon(r)}{r^2}\frac{\left(1+\frac{p(r)}{\varepsilon(r)} \right)\left(
1+\frac{4\pi r^3p(r)}{M(r)}\right)}{\left(1-\frac{2M(r)}{r}\right)} \ ,
\label{e:tov3}
\end{equation}
we can read explicitly the Newtonian part and arrive to an interpretation of the equations that is quite instructive. Think of a shell of matter of radius $r$ and thickness $dr$. Equation (\ref{e:tov2}) gives the mass-energy in this shell. The pressure of the matter exterior to this shell is $p(r)$ and the interior to it $p(r)+dp(r)$. The left side of Eq.\ (\ref{e:tov3}) is the net force acting outward on the surface of the shell by the pressure difference $dp(r)$ and the first factor on the right is the attractive Newtonian force of gravity acting on the shell by the mass interior to it. The remaining factor on the right side of the equation is the exact correction for general relativity. So these equations express the balance at each $r$ between the internal pressure as it supports the overlying material againsts the gravitational attraction of the mass-energy interior to $r$. They are just the equations of hydrostatic equilibrium in general relativity.

The equation of state, {\it i.e.,} the relation between the pressure $p$ and the energy density $\varepsilon$, is the manner in which matter enters the equations of stellar structure. For a given EoS, the TOV equations can be integrated from the origin with the boundary conditions $M(r)=0$ and $p(0)=p_c$, being $p_c$ an arbitrary value,  until the pressure becomes zero. Zero pressure can support no overlying material against the gravitational attraction exerted on it from the mass within and, therefore, marks the edge of the star. The point $R$, where the pressure vanishes, defines the radius of the spherically symmetric star. The integration of Eq.\ (\ref{e:tov2}) from zero up to this value $R$ gives its gravitational mass
\begin{equation}
M_G=4\pi \int_{0}^{R}drr^2\varepsilon(r) \ .
\label{e:mg}
\end{equation}

\subsubsection{Rotating Neutron Stars: Hartle--Thorne approach}

Neutron stars are, however, rotating objects. We expect the rotation to flatten the star more or less depending on its angular velocity $\Omega$. Spherical symmetry is thereby broken although the star maintains its axial symmetry. This symmetry breaking makes the structure equations of rotating neutron stars much more complicated than those of the non-rotating ones. The difficulties which make the construction of models of rotating neutron stars rather cumbersome are due to: (i) the deformation of the star; (ii) the increase of its mass because of the rotation with the consequent modification of the geometry of space-time; and (iii) the general relativistic effect of the dragging of local inertial frames, sometimes referred to as Lense--Thirring effect. 

The properties of rotating neutron stars can be determined by a direct numerical integration of  Einstein's equations \cite{salgado94}.  However, a usual approach followed by many authors to address this problem has been based on a perturbative method developed first by Hartle and Thorne \cite{hartle67,thorne68}. This method was initially though to be valid only for angular velocities much more smaller than the so-called Keplerian angular velocity $\Omega_K$ above which matter is ejected from the star's equator. However, nowadays it is known that, in fact, the method is valid also for angular velocities close to $\Omega_K$ within a few percent of the exact numerical solution. In the following, we present the set of equations that one has to solve, in addition to Eqs.\ (\ref{e:tov1}) and (\ref{e:tov2}), to determine the structure of a rotating neutron star within the Hartle--Thorne approach and that account for the three basic features of a rotating relativitic star: frame dragging, mass increase and rotational deformation. The interested reader is referred to the original work of Hartle and Thorne for a detailed derivation of these equations and a comprehensive discussion (see also, {\it e.g.,} Chapter 15 of Ref.\ \cite{weber} and Chapter 6 of Ref.\ \cite{glendenning}).

The first of these equations accounts for the dragging of the local inertial frame and it allows to determine the difference $\bar{\omega}(r,\theta)=\Omega-\omega(r,\theta)$ between the angular velocity of the star $\Omega$ and that of the local inertial frame $\omega(r)$ located at the position $(r,\theta)$ of a fluid element. It can be shown (see  Ref.\ \cite{hartle67}) that $\bar{\omega}$ is in fact function only of the radial coordinate $r$ and it obeys the differential equation
\begin{equation}
\frac{1}{r^4}\frac{d}{dr}\left(r^4j(r)\frac{d\bar{\omega}(r)}{dr}\right)+\frac{4}{r}\frac{dj(r)}{dr}\bar{\omega}(r)=0 \ , 
\label{e:barom}
\end{equation}
subject to the boundary conditions $\bar{\omega}(0)=\bar{\omega}_c$ and $\left(d{\bar{\omega}(r)}/dr\right)_{r=0}=0$ at the center of the star with $\bar{\omega}_c$ an arbitrary constant value. The function $j(r)$ is defined as
\begin{equation}
j(r) \equiv e^{-\nu(r)}\sqrt{1-\frac{2M(r)}{r}} \ ,
\label{e:jr}
\end{equation}
where the metric function $\nu(r)$ fulfils the equation
\begin{equation}
\frac{d\nu(r)}{dr}=\frac{M(r)+4\pi r^3 p(r)}{r\left(r+2M(r)\right)} \ ,
\label{e:phi}
\end{equation}
with $M(r)$ and $p(r)$ being the solutions of the TOV equations. One can start the integration of this equation with any convenient value of $\nu(0)$, say zero. Outside the star one has
\begin{equation}
\bar{\omega}(r)=\Omega-\frac{2}{r^3}J(\Omega) \ , 
\label{e:barom2}
\end{equation}
where $J(\Omega)$ can be identified with the total angular momentum of the star given by
\begin{equation}
J(\Omega)=\frac{8\pi}{3}\int_{0}^R dr r^4 \frac{p(r)+\varepsilon(r)}{\sqrt{1-\frac{2M(r)}{r}}}\left(\Omega-\omega(r)\right)e^{-\nu(r)} \ ,
\label{e:jang}
\end{equation}
from which the moment of inertia of the star, defined as 
\begin{equation}
I\equiv \frac{J(\Omega)}{\Omega} \ ,
\label{e:momin}
\end{equation}
can be obtained. Note that the relativistic corrections in the moment of inertia come from the dragging of local inertial frames ($\bar{\omega}(r)/\Omega<1$), and the readshift $(e^{-\nu(r)})$ and space curvature $(1/\sqrt{1-2M(r)/r})$ factors.

The increase of mass of the star $\Delta M(\Omega)$ due to the rotation can be obtained from the relation
\begin{equation}
\Delta M(\Omega) = m_0(R) + \frac{J(\Omega)^2}{R^3} \ .
\label{e:deltam}
\end{equation}
In this expression, $R$ is the radius of the non-rotating spherical star and the function $m_0(r)$ is the monopole mass perturbation that can be determined by integrating the equation
\begin{eqnarray}
\frac{dm_0(r)}{dr}&=&4\pi r^2\frac{d\varepsilon}{d p}\left(\varepsilon(r)+p(r)\right)p_0(r)
+\frac{1}{12}j(r)^2r^4\left(\frac{d\bar{\omega}(r)}{dr}\right)^2 \\
&+&\frac{8\pi}{3}r^5j(r)^2\frac{\varepsilon(r)+p(r)}{r-2M(r)}\bar{\omega}(r)^2 
\label{e:m0}
\end{eqnarray}
simultaneously with that for the monopole pressure perturbation $p_0(r)$:
 \begin{eqnarray}
\frac{dp_0(r)}{dr}&=&-\frac{1+8\pi r^2p(r)}{r^2\left(r-2M(r)\right)^2}m_0(r)
-4\pi\frac{\left(p(r)+\varepsilon(r)\right)r^2}{r-2M(r)}p_0(r) \\
&+&\frac{1}{12}\,\frac{r^4j(r)^2}{r-2M(r)}\left(\frac{d\bar{\omega}(r)}{dr}\right)^2
+\frac{1}{3}\frac{d}{dr}\left(\frac{r^3j(r)^2\bar{\omega}(r)^2}{r-2M(r)}\right) 
\label{e:p0} 
\end{eqnarray} 
using as boundary conditions $m_0(0)=p_0(0)=0$.

Finally, the deformation of the star due to the rotation can be characterized in terms of the so-called eccentricity which describes the shape of the star at its surface. It is defined as \cite{fip86}
\begin{equation}
e=\sqrt{1-\left(\frac{R_p}{R_{eq}}\right)^2} \ ,
\label{e:ecc}
\end{equation}
where $R_p$ and $R_{eq}$ are the polar and equatorial radii of the rotationally deformed star which in the
Hartle--Thorne approach are given  by
\begin{eqnarray}
R_{p} \approx R+\xi_0(R)-\frac{1}{2}\xi_2(R)  \ , \,\,\,
R_{eq} \approx R+\xi_0(R)+\xi_2(R)  \ .
\label{e:epradii}
\end{eqnarray}
The quantities $\xi_0(r)$ and $\xi_2(r)$ are the so-called spherical and quadrupole stretching functions:
\begin{eqnarray}
\xi_0(r)&=& -p_0(r)\left(\varepsilon(r)+p(r)\right)\left(\frac{dp(r)}{dr}\right)^{-1}  \\
\xi_2(r)&=& -p_2(r)\left(\varepsilon(r)+p(r)\right)\left(\frac{dp(r)}{dr}\right)^{-1} 
\label{e:xi0}
\end{eqnarray}
defined in terms of $\varepsilon(r)$, $p(r)$, $p_0(r)$ and the quadrupole pressure perturbation $p_2(r)$ given by 
\begin{equation}
p_2(r)=-h_2(r)-\frac{1}{3}\left(r\bar{\omega}(r)\right)^2e^{-2\nu(r)} \ ,
\label{e:p2}
\end{equation}
where the function $h_2(r)$ is solution of the equation
\begin{eqnarray}
\frac{dh_2(r)}{dr}&=&\left(-2\frac{d\nu(r)}{dr}+\frac{2r}{r-2M(r)}\left(\frac{d\nu(r)}{dr}\right)^{-1}\left(2\pi\left(p(r)+\varepsilon(r)\right)-\frac{M(r)}{r^3}\right) \right)h_2(r) \nonumber \\
&-&\frac{2}{r\left(r-2M(r)\right)}\left(\frac{d\nu(r)}{dr}\right)^{-1}v_2(r) \\
&+&\frac{1}{6}\left(r\frac{d\nu(r)}{dr}-\frac{1}{2\left(r-2M(r)\right)}\left(\frac{d\nu(r)}{dr}\right)^{-1} \right)r^3j(r)^2\left(\frac{d\bar{\omega}(r)}{dr}\right)^2 \nonumber \\
&-&\frac{1}{3}\left(r\frac{d\nu(r)}{dr}+\frac{1}{2\left(r-2M(r)\right)} \right)\left(\frac{d\nu(r)}{dr}\right)^{-1}\left(r\bar{\omega}(r)\right)^2\frac{dj(r)^2}{dr}
\label{e:h2}
\end{eqnarray}
which must be  simultaneously integrated together with the equation
\begin{equation}
\frac{dv_2(r)}{dr}=-2\frac{d\nu(r)}{dr}h_2(r)+\left(\frac{1}{r}+\frac{d\nu(r)}{dr}\right)
\left(-\frac{r^3}{3}\frac{dj(r)^2}{dr}\bar{\omega}(r)^2
+\frac{j(r)^2}{6}r^4\left(\frac{d\bar{\omega}(r)}{dr}\right)^2
\right) \ ,
\label{e:v2}
\end{equation}
with boundary conditions $v_2(0)=h_2(0)=0$ and $v_2(\infty)=h_2(\infty)=0$.

In summary, to determine the structure of a rotating neutron star within the Hartle--Thorne approach one must: (i) solve the TOV equations (Eqs.\ (\ref{e:tov1}) and (\ref{e:tov2})) to obtain the mass $M(r)$ and pressure $p(r)$ functions of the non-rotating spherical star; (ii) integrate Eqs.\ (\ref{e:barom}) and (\ref{e:phi}) to compute $\bar{\omega}(r)$ and $\nu(r)$ from which one can evaluate the total angular momentum $J(\Omega)$ of the star and its moment of inertia $I$; (iii) solve Eqs.\ (\ref{e:m0}) and (\ref{e:p0}) to find $m_0(r)$ and $p_0(r)$ and calculate the increase of mass $\Delta M(\Omega)$ due to the rotation, and the spherical stretching function $\xi_0(r)$; and (iv) solve Eqs.\ (\ref{e:h2}) and (\ref{e:v2}) to determine $p_2(r)$, and from it the quadrupole stretching function $\xi_2(r)$ which, together with $\xi_0(r)$, allow to calculate the polar and equatorial radii and the eccentricity $e$ of the star that characterizes its rotational deformation.


\subsection{The nuclear equation of state}
\label{sec:EoS}

\subsubsection{Generalities}

The only ingredient needed to solve the structure equations of neutron stars is the equation of state of dense matter.
Its determination, however, is very challenging due to the wide range of densities, temperatures and isospin asymmetries found in these objects, and it constitutes nowadays one of the main problems in nuclear astrophysics. 
The main difficulties are associated to our lack of a precise knowledge of the behavior of the in-medium nuclear interaction, and to the very complicated resolution of the so-called nuclear many body problem \cite{mbp}. 

Models of the EoS in the neutron star crust are based on reliable experimental data on atomic nuclei, nucleon scattering, and the theory of strongly coupled Coulomb systems. The atomic nuclei in the outer crust are expected to be those studied in the laboratory with a maximum isospin asymmetry $\beta=(N-Z)/A\simeq 0.3$, $N$, $Z$ and $A$ being, respectively, the neutron, proton and total mass number of an atomic nucleus. In the inner crust, as explained in Sec. \ref{sec:anatomy}, nuclei are very neutron rich. Such nuclei, however, do not exist in laboratory because they are beyond the neutron drip line under terrestial conditions. Consequently, our knowledge of the properties of matter under the  density and isospin asymetry conditons characteristic for the inner crust ($10^{11}\leq \rho \leq 10^{14}$ g/cm$^3$ and $0.3 \leq \beta \leq 0.8$) relies on theoretical models.

As seen in Sec.\ref{sec:anatomy} at densities $\sim 10^{14}$ g/cm$^3$ matter becomes a uniform quantum fluid of neutron, protons and electrons. The EoS in the outer core of the neutron star can be calculated in a rather reliable way using models and methods of the nuclear many-body theory which have been applied with some success for the microscopic description of ordinary nuclear structure. However, the reliability of these models and methods decreases when density increases and one enters the inner core region where the true composition of matter is unknown. Theoretical calculations of the nuclear EoS at such extreme densities can be tested exclusively by neutron star observations.

\subsubsection{Experimental and observational constraints of the nuclear EoS}

Properties of nuclear matter can be characterized by a set of few isoscalar ($E_0, K_0, Q_0$) and isovector ($ S, L, K_{sym},Q_{sym}$) parameters that are related to the coefficients of a Taylor expansion of the energy per particle of asymmetric nuclear matter around $\rho_0$ and $\beta=0$  
\begin{equation}
\frac{E}{A}(\rho,\beta)=E_0+\frac{1}{2}K_0x+\frac{1}{6}Q_0x^3+\left(S_0+Lx+\frac{1}{2}K_{sym}+\frac{1}{6}Q_{sym}x^3\right)\beta^2+{\cal O}(4) \ .
\label{ec:expansion}
\end{equation}
Here $x=(\rho-\rho_0)/3\rho_0$, $E_0$ is the energy per particle of symmetric nuclear matter at $\rho_0$, $K_0$ the incompressibility parameter, $Q_0$ the so-called skewness, $S_0$ the value of the nuclear symmetry energy at $\rho_0$, $L$ the slope of the symmetry energy, $K_{sym}$ the symmetry incompressibility, and $Q_{sym}$ the third derivative of the symmetry energy with respect to the density. 

These parameters can be constrained by nuclear experiments. Measurements of nuclear masses yield $E_0=-16\pm 1$ MeV \cite{angeli,wang}. The value of $K_0$ can be extracted from the analysis of isoscalar giant monopole resonances in heavy nuclei. Results of Ref.\ \cite{colo04} suggest  $K_0=240\pm 10$ MeV whereas in Ref.\ \cite{piekarewicz04} a value of $K=248 \pm 8$ MeV is reported while in Ref.\ \cite{blaizot80} it is given $K=210\pm 30$ MeV. Recently, Khan {\it et al.} \cite{elias12} have shown that 
the third derivative $M$ of the energy per unit volume of symmetric nuclear matter is constrained by giant monopone resonance measurements not at $\rho_0$ but rather around what has been called crossing density $\rho\approx 0.11$ fm$^{-3}$. These authors found $M=1100\pm 70$ MeV, whose extrapolation at $\rho_0$ gives $K_0=230\pm 40$ MeV. The value of the skewness parameter $Q_0$ is more uncertain and is not very well constrained yet, being the estimated value in the range $-500\leq Q_0\leq 300$ MeV.

Experimental information on the isovector parameters of the nuclear EoS can be obtained from several sources such as
the analysis of giant \cite{giant} and pygmy \cite{pygmy1,pygmy2} resonances, isospin difussion measurements \cite{isodif}, isobaric analog states \cite{isob}, isoscaling \cite{isoscaling}, measurements of the neutron skin thickness in heavy nuclei \cite{skin1,skin2,skin3,skin4,skin5,skin6,skin7} or meson production in heavy ion collisions \cite{meson1,meson2}. However, whereas $S_0$ is more or less well established ($\sim 30$ MeV), the values of $L$, and specially those of $K_{sym}$ and $Q_{sym}$, are still very uncertain and poorly constrained.  Why the isovector part of the nuclear EoS is so uncertain is still and open question whose answer is related to our limited knowledge of the nuclear force and, in particular, to its spin and isospin dependence.

As mentioned in Sec.\ \ref{sec:anatomy} the presence of other degrees of freedon in addition to nucleons and leptons is expected in the inner core of neutron stars. However, our knowledge of the EoS of such exotic matter is even more uncertain. The main problem in understanding, for instance, the properties of hyperonic matter is the fact that the hyperon-nucleon and hyperon-hyperon interactions are still poorly constrained due to the limited number of experimental data \cite{gal16}. In the case of deconfined quark matter, current theoretical descriptions rely on phenomenological models which are constrained using the few available experimental information on high density matter obtained from heavy-ion collisions.  

Additional and complementary information on the nuclear EoS can be extracted from the observation of neutron stars. Nowadays, the most precise and stringent neutron star constraint on the nuclear EoS comes from the recent determination of the unusually high masses of the millisecond pulsars PSR J1614-2230 \cite{demorest} and PSR J0348+0432 \cite{antoniadis}. These two measurements imply that any reliable model for the nuclear EoS should predict maximum masses at least larger than $2 M_\odot$. This observational constraint rules out many of the existent EoS models with exotic degrees of freedom (particularly those with hyperons), although their presence in the neutron star interior is, however, energetically favorable. This has lead to puzzles like the``hyperon puzzle" \cite{vidanaepja} or the ``$\Delta$" puzzle \cite{drago14b} whose solutions are not easy and presenty are subject of very active research.

As the reader can imagine the simultaneous measurement of both mass and radius of the same neutron star would provide the most definite observational constraint on the nuclear EoS. Unfortunately, althought it is very much desiderable, such a measurement does not exists yet. Further constraints can be obtained from the observational
data on neutron star cooling, measurements of the neutron star moment of inertia or, since very recent, from gravitational wave radiation. Severe constraints on the isovector part of the nuclear EoS can be derived, in particular, from the characterization of the core-crust transition \cite{cc1,cc2,cc3,cc4,cc5,cc6}, the analysis of power-law correlations such as the relation between the radius of the star and the EoS \cite{corre} or the study of oscillations modes such as the r-mode \cite{r1,r2,r3}.

\subsubsection{Theoretical approaches of the nuclear EoS}

Theoretically the nuclear EoS has been determined by many authors using both phenomenological and microscopic many-body approaches. Phenomenological approaches, either nonrelativistic or relativistic, are based on effective interactions that are frequently built to reproduce the properties of nuclei \cite{stone}. Skyrme interactions \cite{skyrme59,skyrmea,skyrmeb} and relativistic mean-field models \cite{rmfa,rmfb} are among the most used ones. Many of such interactions are built to describe nuclear systems close to the isospin symmetric case and, therefore, predictions at high isospin asymmetries should be taken with care. Most Skyrme forces are, by construction, well behaved close to  $\rho_0$ and moderate values of the isospin asymmetry. However,  only certain combinations of the parameters of these forces are well determined experimentally. As a consequence, there exists a large proliferation of different Skyrme interactions that produce a similar EoS for symmetric nuclear mattter but predict a very different one for pure neutron matter. Few years ago, Stone {\it et al.} \cite{stoneb} made an extensive and sistematical test of the capabilities of almost 90 existing Skyrme forces to provide good neutron star candidates, finding that only 27 of these forces passed the restrictive tests imposed. A more stringent constraint has been recenty done by Durtra {\it et al.} \cite{dutra} who have examined the suitability of 240 Skyrme interactons with respect to 11 constraints derived from experimental data and the empitical properties of symmetric matter at and close to saturation. These authors found that only 5 of the 240 analyzed satisfied all the constraints imposed. 

Relativistic mean-field models are based on effective Lagrangians densities where the interaction between baryons is described in terms of meson exchanges. The couplings of nucleons with mesons are usually fixed by fitting masses and radii of nuclei and the properties of nuclear bulk matter, whereas those of other baryons, like hyperons, are fixed by symmetry relations and hypernuclear observables.

Microscopic approaches, on other hand, are based on realistic two- and three-body forces that descrive scatterinng data in free space and the properties of the deuteron. These interactions are based on meson-exchange \cite{m1,m2,m3,m4,m5,m6,m7,m8,m9,m10} or, very recently, on chiral perturbation theory \cite{xft1,xft2,xft3,xft4}. To obtain the EoS one has to solve then the complicated many-body problem whose main dificulty lies in the treatment of the repulsive core, which dominates the short-range of the interaction. Different microscopic many-body approaches has been extensively used for the study of the nuclear matter EoS. These include among others: the Brueckner--Bethe--Goldstone \cite{mbp,bbg} and the Dirac--Brueckner--Hartree--Fock \cite{dbhf1,dbhf2,dbhf3} theories, the variational method \cite{var}, the correlated basis function formalism \cite{cbf}, the self-consistent Green's function technique \cite{scgf1,scgf2} or the $V_{\mbox{low}\,k}$ approach \cite{vlowk}. The interested reader is referred to any of the quoted works for details on these approaches.


\subsection{Neutrino emission and cooling of neutron stars}

In addition to the determination of the nuclear EoS, modelling of neutron star cooling has also concentrated part of the effort of theoretitians. The cooling of the newly born hot neutron stars is driven first by the neutrino emission from the interior,
and then by the emission of photons at the surface. Neutrino emission processes can be divided into slow and fast processes depending on whether one or two baryons participate. The simplest possible neutrino emission process is the so-called direct Urca process:
\begin{equation}
n \rightarrow p+l+\bar \nu_l \ , \,\,\, p+l \rightarrow n +\nu_l \ .
\end{equation}
This is a fast mechanism which however, due to momentum conservation, it 
is only possible when the proton fraction exceeds a critical value $x_{DURCA} \sim 11\%$ to $15 \%$ \cite{lattimer}. Other neutrino processes
which lead to medium or slow cooling scenarios, but that are operative at any density and proton fraction, are the so-called modified Urca processes:
\begin{equation}
N+ n \rightarrow N+ p+l+\bar \nu_l \ , \,\,\, N+p+l \rightarrow N+n +\nu_l \ , 
\end{equation}
the bremsstrahlung: 
\begin{equation}
N+N \rightarrow N+N + \nu +\bar \nu \ ,
\end{equation}
or
the Cooper pair formation: 
\begin{equation}
n+n\rightarrow [nn]+\nu+\bar \nu \ , \,\,\, p+p\rightarrow [pp]+\nu+\bar \nu, 
\end{equation}
this last operating only when the
temperature of the star drops below the critical temperature for neutron superfluidity or proton superconductivity. If hyperons are present in the neutron star interior new 
neutrino emission processes, like {\it e.g.,} 
\begin{equation}
Y\rightarrow B+l+\bar\nu_l \ , 
\end{equation}
may occur providing additional fast cooling mechanisms.  Such additional rapid cooling mechanisms, however, can lead to surface temperatures much lower than 
that observed, unless they are suppressed by hyperon pairing gaps. Therefore, the study of hyperon superfluidity becomes of particular interest since it 
could play a key role in the thermal history of
neutron stars. Nevertheless, whereas the presence of superfluid neutrons in the inner crust, and superfluid neutrons together with
superconducting protons in the core is well established and has been the subject of many studies, a quantitative estimation of the
hyperon pairing has not received so much attention, and just few calculations exists in the literature \cite{super,super2,super3,super4,super5,super6,super7}.


\section{Summary and concluding remarks}
\label{sec:summary}

As we said in the introduction, this work is just a very short overview on the physics of neutron stars where we have tried to present the most remarkable observational and theoretical aspects of this field. Our main intention was to catch the attention on this fascinating topic of the new generation of young students and early-stage researches that attended this school, and motivate them to perform, by their own, more detailed studies. If we have achieved this goal we will feel fully rewarded.

The beginning of the twenty first century has been been particularly generous for the physics of neutron stars. Satellite-based telescopes in different frequency bands are providing us an increible wealth of new observational data. New classes of neutron stars have been discovered. The recent observation of gravitational waves originated from the merger of two neutron stars opens, as alreay said, a new era in the observation of neutron stars from which many surprises are expected. Further surprises are also expected from the new generation of ground-based radio telescope arrays. Last but not least, the new generation of exotic beam facilities in France (SPIRAL), Germany (FAIR), Japan (RIKEN, J-PARC), USA (RIA) or the EU (EURISOL) will allow experimental studies of very exotic nuclei which will have a direct impact on the modeling of neutron stars.

The study of neutron stars is probably one of the most  interdisciplinary fields in physics. Nowadays, it is becoming more and more clear that the only way to unveil the mysteries of neutron stars requieres the strong interplay and collaboration of observers with theoretitians from different areas of physics. Only with the common effort of different communities it will be possible to reach a coherent description and understanding of neutron stars. An example of this necessary common effort is the presently running European network {\it ``PHAROS: The multi-messener physics and astrophysics of neutron stars"} \cite{pharos} which is the continuation of the previous network {\it ``NewCompstar: 
Exploring fundamental physics with compact stars"} \cite{newcompstar}. The aim of this network is to bring together the leading experts in astrophysics, nuclear physics and gravitational physics to study neutron stars through an interdiciplinay approach and provide a dedicated training program for a new generation of scientists.


\section*{Acknowledgements}

The author is very grateful to the organizers of the school I. Bombaci, A. Bonaccorso, G. Casini, M. A. Ciocci, J. Dohet--Elary, A. Kievky, D. Logoteta, L. E. Marcucci, V. Rosso and M. Viviani for their kind invitation to give this lecture, and to all the students for their interest and questions. The author wants also to thank D. Unkel for the interesting discussions they had during the preparation of this work. This work has been supported by ``PHAROS: The multi-messener physics and astrophysics of neutron stars", COST Action CA16214. 




\end{document}